\documentclass[%
 reprint,
 amsmath,amssymb,
 aps,
]{revtex4-2}

\usepackage{graphicx}
\usepackage{dcolumn}

\usepackage{algorithmic}
\usepackage{algorithm}
\usepackage[pdfauthor={derajan}, pdftitle={How to do this}, pdfstartview=XYZ, bookmarks=true, colorlinks=true, linkcolor=blue, urlcolor=blue, citecolor=blue, pdftex, bookmarks=true, linktocpage=true, hyperindex=true ]{hyperref}
\usepackage{color}

\newcommand{\syt}[2]{\textcolor{blue}{#2}}
\usepackage[inkscapelatex=false]{svg}
\begin{document}

\preprint{APS/123-QED}

\title{
RBM-Based Simulated Quantum Annealing for Graph Isomorphism Problems
}

\author{Yukun Wang\textsuperscript{1,2}}
 \email{wykun06@gmail.com}
\author{Yingtong Shen\textsuperscript{1,2}}
 \email{yingtong0409@163.com}
 \author{Xinyao Wu}
\author{Zhichao Zhang\textsuperscript{3}}
\author{Linchun Wan\textsuperscript{4} }

\affiliation{\textsuperscript{1} Beijing Key Laboratory of Petroleum Data Mining, China University of Petroleum, Beijing 102249, China}
\affiliation{\textsuperscript{2} State Key Lab of Processors, Institute of Computing Technology, CAS, Beijing 100190, China}
\affiliation{\textsuperscript{3} School of Mathematics and Physics, University of Science and Technology Beijing, Beijing 100083, China}
\affiliation{\textsuperscript{4} School of Computer and Information Science, Southwest University, Chongqing 400715, China}


\date{\today}

\begin{abstract}
The graph isomorphism problem remains a fundamental challenge in computer science, driving the search for efficient decision algorithms. Due to its ambiguous computational complexity, heuristic approaches such as simulated annealing are frequently used, achieving high solution probabilities while avoiding exhaustive enumeration. However, traditional simulated annealing usually struggles with low sampling efficiency and reduced solution-finding probability in complex or large graph problems. In this study, we integrate the principles of quantum technology to address the graph isomorphism problem. By mapping the solution space to a quantum many-body system, we developed a parameterized model for variational simulated annealing. This approach emphasizes the regions of the solution space that are most likely to contain the optimal solution, thereby enhancing the search accuracy.
Artificial neural networks were utilized to parameterize the quantum many-body system, leveraging their capacity for efficient function approximation to perform accurate sampling in the intricate energy landscapes of large graph problems.

\begin{description}
\item[Usage]
Secondary publications and information retrieval purposes.

\end{description}
\end{abstract}

\maketitle


\section{\label{sec:level1}Introduction}


Graph isomorphism is an equivalence relation that groups graphs sharing the same structure into isomorphism classes. Originally emerging in the 1950s when chemists applied it to compare molecular structures \cite{ray1957finding}, this concept has become a fundamental challenge in graph theory and theoretical computer science. The problem centers on determining whether two graphs can be rendered identical through a simple relabeling of vertices while preserving their connectivity. Despite its pivotal role in areas such as pattern recognition \cite{riesen2015structural}, biochemistry\cite{merkys2023graph}, and communication security \cite{lestringant2015automated}, the computational complexity of the problem, particularly the possibility of a polynomial-time solution, remains unresolved.

Graph isomorphism gained attention in computer science in the 1970s, with early work focusing on efficient algorithms for specific graph classes and exploring its computational complexity \cite{mathon1979note}. In the 1980s, Luks\cite{luks1982isomorphism} introduced decomposition-based algorithms, improving the efficiency of solving the graph isomorphism for certain graph types. A breakthrough came in 2015, when Babai developed an algorithm that solved the graph isomorphism in quasi-polynomial time \cite{babai2016graph}. Despite this progress, the classification of the problem as complete in P or NP remains unresolved, leaving it an open question in theoretical computer science \cite{luks1982isomorphism}. The inherent complexity of the graph isomorphism problem makes it challenging to efficiently and accurately determine correct mappings, especially for large or intricate graphs, driving the development of heuristic approaches. Influential classical methods include tree search with backtracking \cite{conte2004thirty}, tree-based recursive decomposition and decision tree-based algorithms \cite{messmer1998new}, Nauty algorithm\cite{mckay1981practical},  LAD\cite{solnon2010alldifferent}, and VF series algorithms that incrementally match vertex pairs \cite{cordella2004sub,cordella2001improved}.
These algorithms often incorporate heuristic strategies such as searching for the minimum of an energy function in vertex/edge correspondence mismatches and combining empirical rules (e.g., vertex neighborhoods, degree sequences, subgraph structures) to prune the search space and expedite approximate solutions. Among these, the Simulated Annealing Algorithm (SA) has emerged as a notable method inspired by thermodynamic processes\cite{zeguendry2023quantum, onizawa2022fast, xiutang2008simulated, delahaye2019simulated}; however, its efficiency is often hindered by slow sampling in rugged optimization landscapes and multiple local extrema, underscoring the ongoing need for novel algorithms.

Quantum computing, leveraging inherent parallelism, holds promise for addressing the graph isomorphism problem. One class of approaches uses quantum walks to encode graph connectivity into qubit probability amplitudes, potentially enabling more efficient vertex matching while reducing storage overhead compared to classical stepwise comparisons \cite{tamascelli2014quantum,gamble2010two,rudinger2012noninteracting,wang2018marking,li2023graph}. 
Alternatively, quantum annealing reformulates the graph isomorphism problem as an unconstrained quadratic binary optimization problem (QUBO), which is then mapped onto an Ising spin glass model \cite{warren2013adapting,minamisawa2019high,hertz1991introduction}. In principle, finding the ground state of the corresponding Hamiltonian yields an optimal solution, and devices such as D-Wave quantum annealers have demonstrated this approach experimentally \cite{qiang2021implementing,wang2016quantum,gaitan2014graph,borowski2020new}. Finally, several schemes directly encode entire graph structures into quantum states to measure invariants that distinguish non-isomorphic pairs, though these generally yield only necessary (not sufficient) conditions and do not fully reconstruct the vertex mapping \cite{mills2019quantum,bradler2021graph,qiang2021implementing}.

Despite the potential advantages of quantum algorithms for graph isomorphism, challenges remain. Currently, quantum computing operates in the noisy intermediate-scale quantum (NISQ) era, where quantum noise and limited qubit number affect the performance of even theoretically mature quantum technologies.   
Algorithmically, quantum walks may be hindered by the regularity of the graph structure, particularly for strongly regular graphs. Quantum annealing requires strict adherence to the adiabatic condition, which is difficult to achieve for large-scale graphs due to complex Hamiltonian structures and small energy gaps. Although NISQ devices hinder the full potential of quantum computing \cite{di2024quantum}, they can still enhance classical algorithms by simulating quantum principles. This has led to the rise of hybrid quantum-classical algorithms\cite{li2023fast, crosson2016simulated, bando2021simulated, zeng2024performance, vargas2021many}, integrating classical machine learning models with quantum computing, which has become a key area of research. One notable approach is Simulated Quantum Annealing (SQA)\cite{inack2015simulated, crosson2016simulated, bando2021simulated}, a quantum-inspired classical algorithm that employs the Path Integral Monte Carlo (PIMC) method\cite{martovnak2002quantum,stella2006monte,isakov2016understanding,mbeng2019dynamics} to map the quantum Hamiltonian onto a classical system via Trotter decomposition (forming “Trotter slices”).  By simulating quantum tunneling effects, SQA offers a higher probability to escape local optima compared to  SA or classical heuristics.


{However, dividing the quantum evolution into Trotter slices sharply increases computational costs with growing slice numbers and system sizes, while using only a finite number of slices truncates quantum fluctuations and introduces approximation errors in the quantum-to-classical mapping. These factors ultimately restrict SQA’s effectiveness for large-scale graph isomorphism problems.  
This work introduces an SQA algorithm built on a neural-network framework. By employing a Restricted Boltzmann Machine (RBM) to parameterize the quantum system, we recast the graph isomorphism problem as a quantum many-body challenge and solve it using variational neural-network quantum states (NQS). 
NQS has garnered significant interest in quantum mechanics for solving problems such as ground states \cite{carleo2017solving}, dynamic evolution \cite{nomura2017restricted}, and quantum tomography \cite{zeguendry2023quantum, carleo2019machine, torlai2018neural}.
Unlike existing PIMC-based SQA methods, our approach constructs an RBM-based effective Hamiltonian that captures the system’s energy landscape. In principle, by dynamically optimizing a small sampling space, our method increases the likelihood of identifying the global optimum and reduces the risk of being trapped in local optima. 
 Moreover, rapid variational optimization can potentially reduce the computational time per iteration, making our approach particularly attractive for large-scale graph problems. 
} 
\syt{#2}{ 
}

This paper is organized as follows. Section $2$ introduces the definition of graph isomorphism problem and maps it to the QUBO problem and the Ising optimization problem\cite{calude2017qubo,yoshimura2021mapping}. Then, in Section $3$ we explain the methods and technical details to solve the graph isomorphism problem. Then, Section $4$ shows the performance and experimental results of our algorithm on graph instances, and compares the performance with traditional annealing algorithms and some necessary numerical analysis. Finally, in Section $5$, we think about the theoretical significance of the algorithm proposed in this paper and the future direction of research.

\section{Preliminaries}\label{Preliminaries}

\subsection{QUBO Formulation for the Graph Isomorphism Problem}
The graph isomorphism problem is a fundamental question in graph theory, involving the possibility of relabeling a graph's vertices to make it structurally identical to another graph. For two graphs $G_1$ and $G_2$ to be isomorphic, they must satisfy two fundamental constraints: the vertex correspondence constraint and the edge invariant constraint. The vertex correspondence constraint means that there exists a one-to-one mapping between the vertex set $V_1$ of graph $G_1$ and the vertex set $V_2$ of graph $G_2$.The edge invariant constraint requires that for every edge in $G_1$, there is a corresponding edge in $G_2$, and the structure of the graph remains unchanged according to the mapping between the vertices. In this context, we assume that the graphs under discussion are simple undirected graphs without multiple edges or self-loops unless otherwise specified. The key to solving the graph isomorphism problem is to find a mapping between the two sets of vertices that satisfies both the vertex correspondence and edge invariant constraints. From the perspective of the adjacency matrix, if $A_1$ and $A_2$  represent the adjacency matrices of graphs  $G_1$ and $G_2$, respectively, then for the graphs to be isomorphic, the vertex mapping that satisfies the two constraints must correspond to a specific permutation matrix $P$:
\begin{equation}
PA_1P^{-1}=A_2.
\label{eq:1}
\end{equation}

\begin{figure}[h]     
\centering  
\includegraphics[width=8cm]{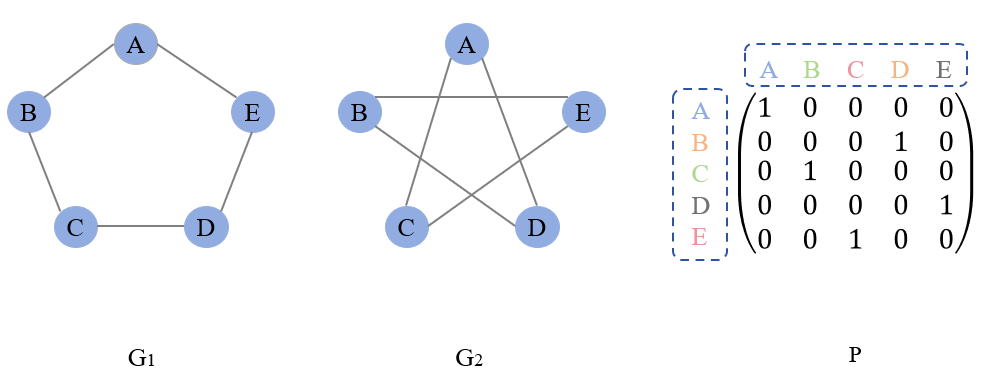}
\caption{$5$-vertex isomorphic graph instance. The matrix $P$ that reflects the mapping relationship between graphs $G_1$ and $G_2$. $P$ is constructed by numbering each vertex and associating each row and column with a node. $1$ is entered into the position that correctly corresponds to the vertex, and $0$ is entered into the rest.} \label{fig1_1}
\end{figure}

As illustrated in Fig.\ref{fig1_1},  graphs $G_1$ and $G_2$ are isomorphic, by mapping $\{A,B,C,D,E\}$ in $G_1$ to $\{ A, C, E, B, D\}$ in $G_2$. The vertex mapping is encoded in the matrix $P$. To transfer the isomorphic problem into a QUBO problem, we then transfer the permutation matrix $P$ into a one-dimensional solution vector:
\begin{equation}
\vec{x} = 
\begin{aligned}
&[x_{AA}, x_{AB}, x_{AC}, x_{AD}, x_{AE}, \\
&x_{BA}, x_{BB}, x_{BC}, x_{BD}, x_{BE}, \\
&x_{CA}, x_{CB}, x_{CC}, x_{CD}, x_{CE}, \\
&x_{DA}, x_{DB}, x_{DC}, x_{DD}, x_{DE}, \\
&x_{EA}, x_{EB}, x_{EC}, x_{ED}, x_{EE}],
\end{aligned}
\label{eq:2}
\end{equation}
where each component $x_{ij} (i\in V_1,j\in V_2)$ is either $0$ or $1$,depending on the mapping relationship.

However, in practice, many graph isomorphism instances involve non-regular graphs, which allows for further optimization of the one-dimensional solution vector $\vec{x}$. As is known,  two graphs with differing degree sequences cannot be isomorphic \cite {quintero2022qubo}. Even when their degree sequences coincide, any valid mapping must pair vertices of identical degree. Exploiting this constraint enables a reduction in the solution space.

\begin{figure}[h]     
\centering  
\includegraphics[width=8cm]{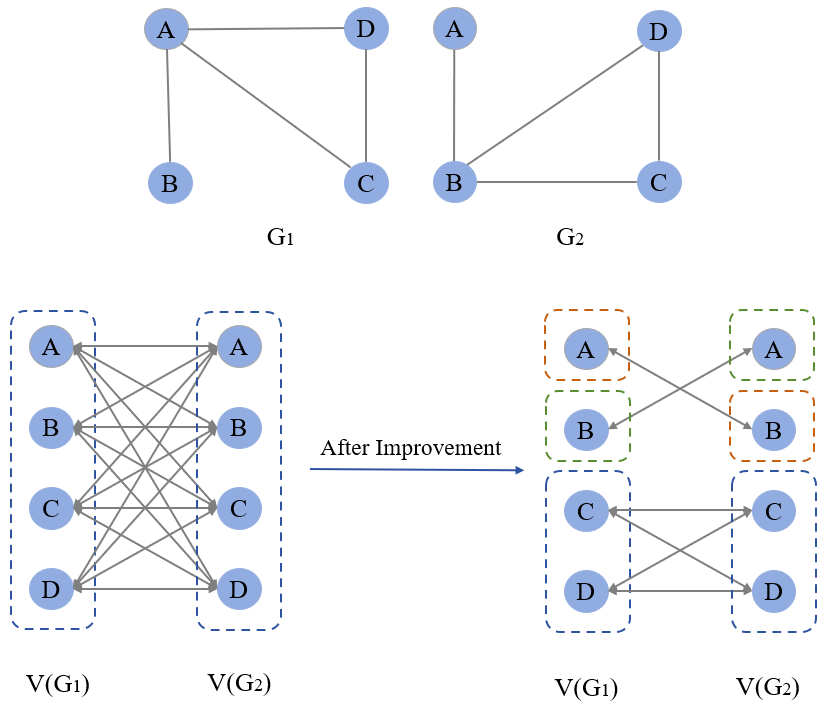}
\caption{Ordinary four-vertex isomorphic graphs. The improved encoding method can shorten the length of the solution vector in some non-regular cases.} \label{fig3}
\end{figure}

To illustrate, consider Fig.\ref{fig3}, where graphs $G_1$ and $G_2$ have the same degree sequence $\{1,2,3,3\}$. The vertex set can be divided into three smaller subsets,  with only vertices of the same degree eligible for mapping to each other. This reduces the solution vector length from $16$ to $6$, as only the following mappings are possible: $B$ to $A$, $\{A,C\}$ to $\{B, D\}$, $D$ to $C$. Thus there has $
\vec{x} = 
[x_{AA}, x_{AB}, x_{AD}, x_{BA}, x_{CB}, x_{CD}, x_{DC}]$. This optimization significantly reduces the solution space and makes the isomorphism problem more tractable.

After encoding the solution space into the vector $\vec{x}$, the graph isomorphism problem can then be reformulated as a QUBO problem by defining an objective function $F(\vec{x})$ to be minimized:
\begin{equation}
\min_{x_i \in \{0,1\}} F(\vec{x}) = \vec{x}^TQ\vec{x},
\label{eq:3}
\end{equation}
where $Q$ is a symmetric matrix related to the problem. The matrix element $Q_{ii}$ represents the linear coefficient of a single variable in $\vec{x}$, while $Q_{ij}$ represents the quadratic interaction coefficient between two variables.

$\vec{x}=\{x_{ij}\}^{|S|}$  is a binary solution vector of length $|S|$ and $x_{ij} \in \{0,1\}$represents the mapping relationship between vertices $i \in V_1$ and $j \in V_2$. Let $S=\{(i,j)|i\in V_1,j \in V_2\}$ be the set of possible vertex mapping pairs, considering the consistency of the node degrees between 
$V_1$ and $V_2$. The objective function  $F(\vec{x})$ is designed to enforce the two fundamental constraints of graph isomorphism:
\begin{equation}
F(\vec{x})=F_1(\vec{x})+F_2(\vec{x}),
\label{eq:4}
\end{equation}
where
\begin{equation}
F_1(\vec{x})=\sum_{i \in V_1}(\sum_{(i,j)\in S}x_{ij}-1)^2+\sum_{i \in V_2}(\sum_{(i,j)\in S}x_{ij}-1)^2,
\label{eq:5}
\end{equation}
and
\begin{equation}
F_2(\vec{x})=\sum_{(i,j) \in S}\sum_{(k,l)\in S}[x_{ij}x_{kl}(A_{1(i,k)}-A_{2(j,l)})^2].
\label{eq:6}
\end{equation}

The first part $F_1(\vec{x})$ is used to ensure the properties of the permutation matrix corresponding to $\vec{x}$, that is, each row and each column has only one element that is $1$. This ensures compliance with the vertex correspondence constraint. The second part $F_2(\vec{x})$ ensures that the mapping relationship complies with the second constraint. The term $(A_{1(i,k)}-A_{2(j,l)})^2$ ensures that only edges matching in both graphs contribute to minimizing $F_2(\vec{x})$. We solve the graph isomorphism problem by minimizing this objective function. Let $\vec{x_0}$ be the optimal solution finally obtained. If $F(\vec{x_0})=0$, this implies that $G_1$ and $G_2$ are isomorphic. The binary entries of $\vec{x_0}$ encode the vertex mapping between $G_1$ and $G_2$ satisfying both constraints. If $F(\vec{x_0})>0$ the graphs are not isomorphic, as no valid mapping exists that satisfies both constraints simultaneously.

\subsection{Ising optimization problem}

The Ising model originates from the spin system in physics and was originally used to describe the phase transition behavior of ferromagnetic materials \cite{volkov1974contribution}. One of the goals of studying the Ising model is to observe the change in the total energy of the system under the interaction of spins and the influence of external magnetic fields. In the Ising model, each atom or spin is assumed to be in one of two states, spin up or spin down. These spins are distributed in a lattice and interact with each other's nearest neighbor spins.

The goal of the Ising optimization problem is to evolve an $N$ qubit quantum system $|\psi\rangle$ to the ground state $|\vec{s}\rangle=|s_0,s_1,...,s_{N-1}\rangle$ of the problem Hamiltonian $\hat{H}$, so that the system energy value $\langle\vec{s}|\hat{H}|\vec{s}\rangle$ is minimized. Therefore, the goal of the Ising optimization problem usually coincides with solving for the minimum value of the objective function in the QUBO problem. The difference is that the boolean variables in the QUBO problem correspond to the projection results $\pm 1$ of the single-bit quantum eigenstates $|+\rangle$ and $|-\rangle$ in the quantum Ising model on the Pauli-Z operator.To describe quantum systems, we use the Pauli-Z operator $\sigma_i^z$ to replace every classical binary variable. The Hamiltonian of the quantum Ising model is generally defined as
\begin{equation}
\hat{H}=-\sum_{i,j=1}^NJ_{ij}\sigma_i^z\sigma_j^z-\sum_{i=1}^Nh_i\sigma_i^z,
\label{eq:7}
\end{equation}
where $J_{ij}$ is the coupling constant describing the spin pair $\sigma_i^z$,$\sigma_j^z$, and $h_i$ is the effect of the external magnetic field on each spin.

\section{RBM-Based Simulated Quantum Annealing  for Graph Isomorphism}\label{NQS}
\subsection{Hamiltonian Construction of Graph Isomorphism Problem}

According to the definition of the Ising model, for a graph isomorphism instance with a solution vector $\vec{x}$ of length $L$, $L$ spin/qubits are used for encoding. The binary variable$x_{ij}$is mapped to a spin/qubit $s_{ij}$  via the Pauli-Z operator, which is used to construct the problem Hamiltonian $\hat{H}$. The solution space of the original problem is then mapped to a quantum system$|\psi\rangle$ of size $2^L$, where $(i,j) \in S$, the vertex mapping pair set. Therefore, the ground state of $\hat{H}$ contains vertex mapping relationships, and the corresponding ground state energy $\lambda$ is an important basis for evaluating whether the mapping relationship is correct. From the two constraints of the graph isomorphism problem, we can see that:
\begin{equation}
\hat{H}=\hat{H_1}+\hat{H_2},
\label{eq:8}
\end{equation}
where
\begin{equation}
\hat{H_1}=\sum_{i=1}^N(\sum_{(i,j)\in S}\frac{\sigma^z_{ij}+I}{2}-I)^2+\sum_{j=1}^N(\sum_{(i,j)\in S}\frac{\sigma^z_{ij}+I}{2}-I)^2,
\label{eq:9}
\end{equation}
and
\begin{equation}
\hat{H_2}=\sum_{(i,j)\in S}\sum_{(k,l)\in S}\frac{1}{4}[(\sigma^z_{ij}\sigma^z_{kl}+\sigma^z_{ij}+\sigma^z_{kl}+I)(A_{1(i,k)}-A_{2(j,l)})^2].
\label{eq:10}
\end{equation}

The eigenstate of 
$\hat{H}$ correspond to the different configurations of $x_{ij}$, where each configuration (with $x_{ij}\in\{0,1\}$) represents a possible mapping relationship in the graph isomorphism problem, and is denoted as $|\vec{s}\rangle$. For any eigenstate $|\vec{s}\rangle$ in the quantum system, which serves as the quantum representation of a mapping solution to the graph isomorphism problem, the associated energy eigenvalue is defined as follows:
\begin{equation}
\langle\vec{s}|\hat{H}|\vec{s}\rangle=\begin{cases}
0, & \text {the right mapping combination} \\
\lambda, &\text{otherwise}
\end{cases}
\label{eq:11}
\end{equation}

Here, $\lambda$ records the number of configuration items that violate the constraints of the graph isomorphism mapping. Its value is determined by equations \eqref{eq:5} and \eqref{eq:6}.Initially, the bijection-related objective function in equation \eqref{eq:5} is evaluated. Notably, any redundant mapping increases the value of $F_1(\vec{x})$  by $1$. In order to determine the upper bound of $F_1(\vec{x})$, it is essential to include the maximum possible number of redundant mappings beyond the valid bijection that satisfies the cost function. Consider two simple undirected graphs, $G_1$ and $G_2$ each with $N$ vertices. Assume that every vertex in $G_1$ maps to every vertex in $G_2$, and each vertex has one correct mapping among the $N$ mapping relationships, with the remaining mappings being redundant. According to equation \eqref{eq:5}, this results in a penalty value of $2N(N-1)^2$.It is important to note that when traversing the vertices of $G_1$ and $G_2$ accounts for  $F_1(\vec{x})$each redundant mapping twice. Therefore, the upper bound of  $F_1(\vec{x})$ is given by $2N(N-1)^2$.

Then, we check the edges mapping objective function $F_2(\vec{x})$. Under the condition that $F_1(\vec{x})$ reaches its maximum value,  the length of $\vec{x}$ is $N^2$, implying that both graphs $G_1$ and $G_2$ are fully connected.As a result, $x_{ij}=x_{kl}=1,\forall (i,j),(k,l)\in S$. For the term $(A_{1(i,k)}-A_{2(j,l)})^2$ , the following cases are considered:  
\begin{equation}
(A_{1(ik)}-A_{2(jl)})^2=\begin{cases}
0,&\text{if $i \ne k$ and $j \ne l$},\\
0, & \text {if $i=k$ and $j=l$ }, \\
1, &\text{others},
\end{cases}
\label{eq:12}
\end{equation}
when $i\ne k$ and $j\ne l$, means under the mapping $x_{ij}=x_{kl}=1$, the edge invariant constraint.In the case  $i= k$ and $j= l$, there is $A_{1(i,k)}=A_{2(j,l)}=0$
since there is no cycle in a simple undirected graph. The third case, where $i=k$ and $j \ne l$ or $i \ne k$ and $j=l$, the absence of cycles in the graph ensures that $A_{1(i,k)}$ and $A_{2(j,l)}$ are not simultaneously zero, resulting in the theoretical maximum value $2N^2(N-1)$. $F_{2}(\vec{x})$
checks whether the edge relationships are consistent for mapped vertex pairs, incrementing its value by 1 for each violation. However, it does not verify compliance with bijection principles, as this is constrained by $F_1(\vec{x})$. Thus, the maximum value of equation \eqref{eq:5} is added to the maximum value of equation \eqref{eq:6}, yielding an upper bound for $\lambda$ of $2N^2(N-1)+2N(N-1)^2$.

\subsection{Neural Network Quantum States Approximate Wavefunctions}

Above, we have encoded the graph isomorphism problem into the Hamiltonian $\hat{H}$, where the lowest energy state of 
$\hat{H}$ captures the isomorphism information of the graph. In principle, we could continue to exploit quantum advantages through methods such as quantum annealing, QAOA, or quantum walks to solve the problem. However, considering the current limitations of practical quantum devices, algorithms that employ quantum-inspired classical simulations can also offer advantages. Therefore, we focus on applying quantum-inspired approaches to solve the problem via classical simulation.
While the SQA algorithm has been widely researched, it has certain limitations. To address these issues, we use NQS to represent the solution space. We then apply the quantum Monte Carlo method to estimate the ground state energy as accurately as possible, which helps solve these challenges.

NQS is an innovative method for representing quantum states\cite{jia2019quantum,carleo2017solving}, widely used in quantum computing and quantum physics. In quantum physics, the many-body problem refers to describing the behavior of a system composed of a large number of interacting particles. The wave function of such a system has exponential complexity because the interaction between each particle and all other particles needs to be considered \cite{verstraete2015worth,osborne2012hamiltonian}. This results in analytical difficulties in describing and solving the wave function for complex quantum systems. Following the research of \cite{carleo2017solving}, the strong expressiveness of neural network models in machine learning provides an approximate solution to this problem, which has also greatly promoted the research progress of condensed matter physics\cite{jia2019efficient,gao2017efficient,carrasquilla2017machine,monras2017inductive,dunjko2016quantum}.

In this work, we first choose the more flexible and simpler neural network model, the Restricted Boltzmann Machine (RBM), to parameterize the quantum system's wave function. In theory, the wave function of a multi-body quantum system requires an exponential amount of data to fully encode. However, in practical applications, we hope to select a suitable representation method to appropriately simplify the wave function of the system related to the problem, and the graph model is good at this. Then we perform the quantum Monte Carlo method to obtain the ground state energy as accurately as possible. Of course, the ultimate goal of the graph isomorphism problem is not only to obtain the energy value, which serves as the basis for determining whether two graphs are isomorphic, but also to increase the probability that the target candidate mapping solution is measured by evolving the system to the ground state. In our work, the measurement process for simulating the quantum system is replaced by sampling.

\begin{figure}[h]     
    \centering     
    \includegraphics[width=8.5cm]{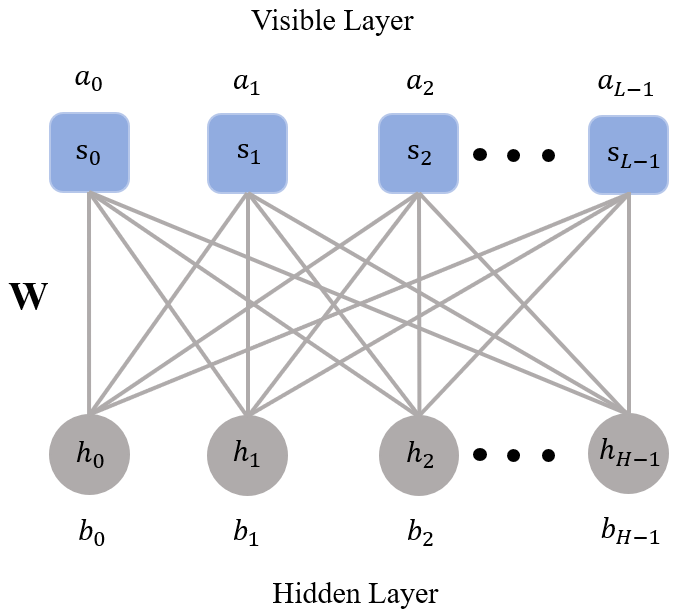}  
    \caption{Structure of a Restricted Boltzmann Machine.}     
    \label{fig5} 
\end{figure}

RBM is a two-layer bipartite graph \cite{zhang2018overview}, typically consisting of two types of units: hidden layer units and visible layer units. RBM use hidden units to model high-order and non-linear patterns in data, enhancing the capability to represent wave functions \cite{torlai2018latent}. The structure of the model is shown in Fig.\ref{fig5}. For a solution vector $\vec{x}$ of length $L$ in the graph isomorphic problem, the corresponding quantum eigenstate is defined as $|\vec{s}\rangle={|s_i\rangle}^{\otimes L}$, which is used as input to interact with the bias vector$\vec{a}=\{a_i\}$ in the visible layer. This input is fully connected to the hidden layer containing  $H$ binary hidden nodes. The bias vector of the hidden layer is $\vec{b}=\{b_j\}$,where $s_i,h_j\in\{1,-1\}$. The connection relationship between each vertex in the two layers is determined by the corresponding connection weight parameter $W=\{W_{ij}\}$. The complete set of parameters for the RBM model is given by $\Omega=\{\vec{a},\vec{b},W\}$, where $\Omega_i \in R$. Thus, the wave function of the quantum system can be expressed as:
\begin{equation}
\psi(\vec{s},\Omega)=\sum_{\vec{h}}e^{\sum_i s_ia_i+\sum_{ij}s_iW_{ij}h_j+\sum_jh_jb_j}
\label{eq:13}
\end{equation}
Given that the units in the RBM layers are not connected within their respective layers, we can simplify this expression to:
\begin{equation}
\psi(\vec{s},\Omega)=e^{\sum_ia_is_i}\prod_j2\cosh(b_j+\sum_iW_{ij}s_i)
\label{eq:14}
\end{equation}
Therefore, after mapping the solution space of the graph isomorphism problem to the quantum system, the corresponding neural network quantum state \cite{zyczkowski2011generating}can be written as:
\begin{equation}
|\psi \rangle = \sum_{\vec{s}}\psi(\vec{s},\Omega)|\vec{s}\rangle
\label{eq:15}
\end{equation}
Now, to calculate the expectation value of the Hamiltonian $\hat{H}$,we use,
\begin{equation}
\langle\psi(\vec{s},\Omega)|\hat{H}|\psi(\vec{s},\Omega)\rangle= |\psi(\vec{s},\Omega)|^2\lambda=P(\vec{s})\lambda
\label{eq:16}
\end{equation}

Here,$|\psi(\vec{s},\Omega)\rangle = \psi(\vec{s},\Omega)|\vec{s}\rangle$  used to represent the state of non-normalized mapping $|\vec{s}\rangle$. Therefore, we realize the parameterized representation of the quantum system through ansatz in equation\eqref{eq:14}. In quantum mechanics, the eigenstate wave function in the system determines the probability that the eigenstate is observed to some extent, as shown in equation\eqref{eq:15}. This provides a theoretical foundation for mapping quantum systems into probability distribution space.

Next, we will apply an optimization program to perform variational operations on the system’s wave function. The goal is to solve for the ground state of the system and obtain the optimal mapping of the graph vertices.

\subsection{Variational Solution of Graph Isomorphism Problem using NQS}
\begin{figure}[h]     
    \centering     
    \includegraphics[width=8.5cm]{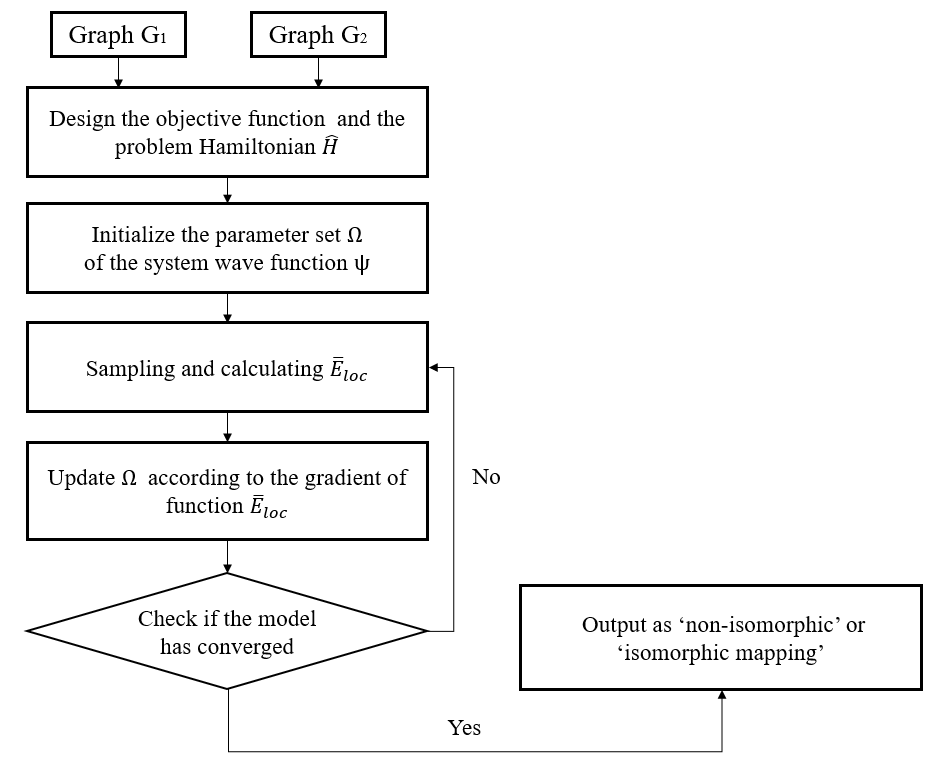} 
    \caption{Proposed RBM-based GI algorithm.}    
    \label{fig11} 
\end{figure}

In this work, we integrate the variational wave function obtained from the neural network model into the variational Monte Carlo (VMC) method. By adjusting the energy of the quantum system through Monte Carlo sampling and applying the variational principle, we aim to approximate the ground state and minimize the energy of the system. In detail, for the problem Hamiltonian $\hat{H}$ of the graph isomorphism problem,  the energy of the system represented by the NQS state $|\psi\rangle$ is given by:
\begin{equation}
E_{global}=\frac{\langle \psi|\hat H|\psi\rangle}{\langle \psi|\psi\rangle}
\label{eq:17}
\end{equation}
This represents the expectation value of the Hamiltonian $\hat{H}$ with respect to the NQS $|\psi\rangle$, providing a measure of the system's energy. 

Next, we assume that $n$ samples $\{|\vec{s}\rangle\}^n$, representing a set of eigenstates of $\hat{H}$, are collected from the system using the Metropolis-Hastings method. The Metropolis-Hastings algorithm generates a sequence of samples from the target distribution $P(\vec{s})$ which means the modulus squared distribution of the system's wave function. by constructing a Markov chain. It accepts or rejects new samples based on the relative values of the target probability distribution at the current and proposed states \cite{metropolis1953equation}.
\begin{equation}
E_{loc}=\sum_{\vec{s}}\langle\psi(\vec{s},\Omega)|\hat{H}|\psi(\vec{s},\Omega)\rangle=\sum_{\vec{s}}P(\vec{s})\lambda
\label{eq:18}
\end{equation}
\begin{equation}
\bar E_{loc}=\frac{1}{M}\sum^ME_{loc }\approx E_{global}
\label{eq:19}
\end{equation}

\begin{figure*}     
    \centering     
    \includegraphics[width=5.0in]{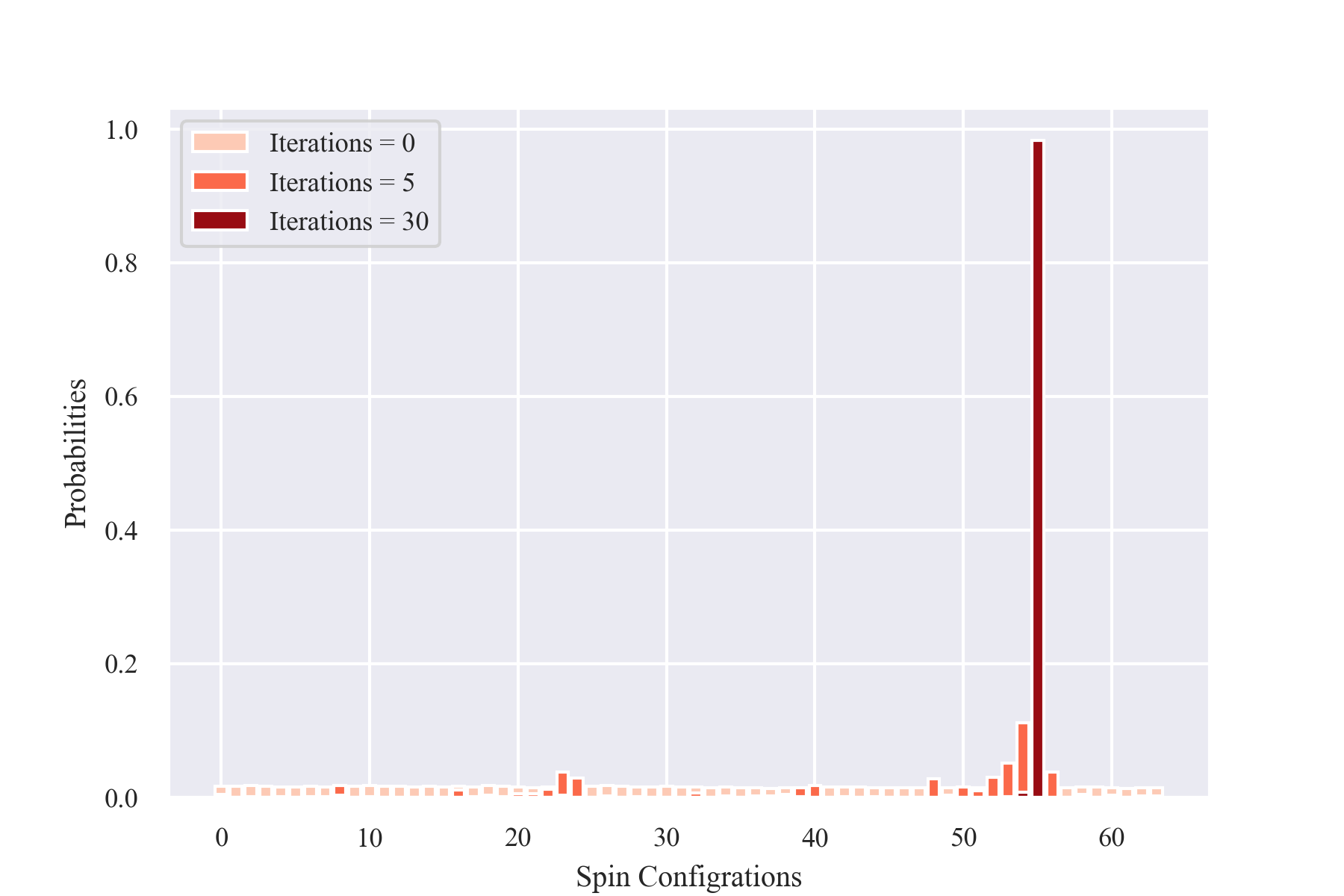} 
    \caption{Schematic of the sampling‐optimization process for the example in Fig.\ref{fig3}. The encoding qubit string is $\vec{x}=[x_{AB},x_{BA},x_{CC},x_{CD},x_{DC},x_{DD}]$, yielding a solution space of size $2^6$.  We perform 30 optimization iterations,  each generating three samples.  It can be seen that the NQS optimization process quickly increases the probability of sampling high-quality solutions (which means producing lower penalty function values), thereby making the target solution stand out.}    
    \label{fig6} 
\end{figure*}

\begin{figure*}     
\centering  
\includegraphics[width=4.0in]{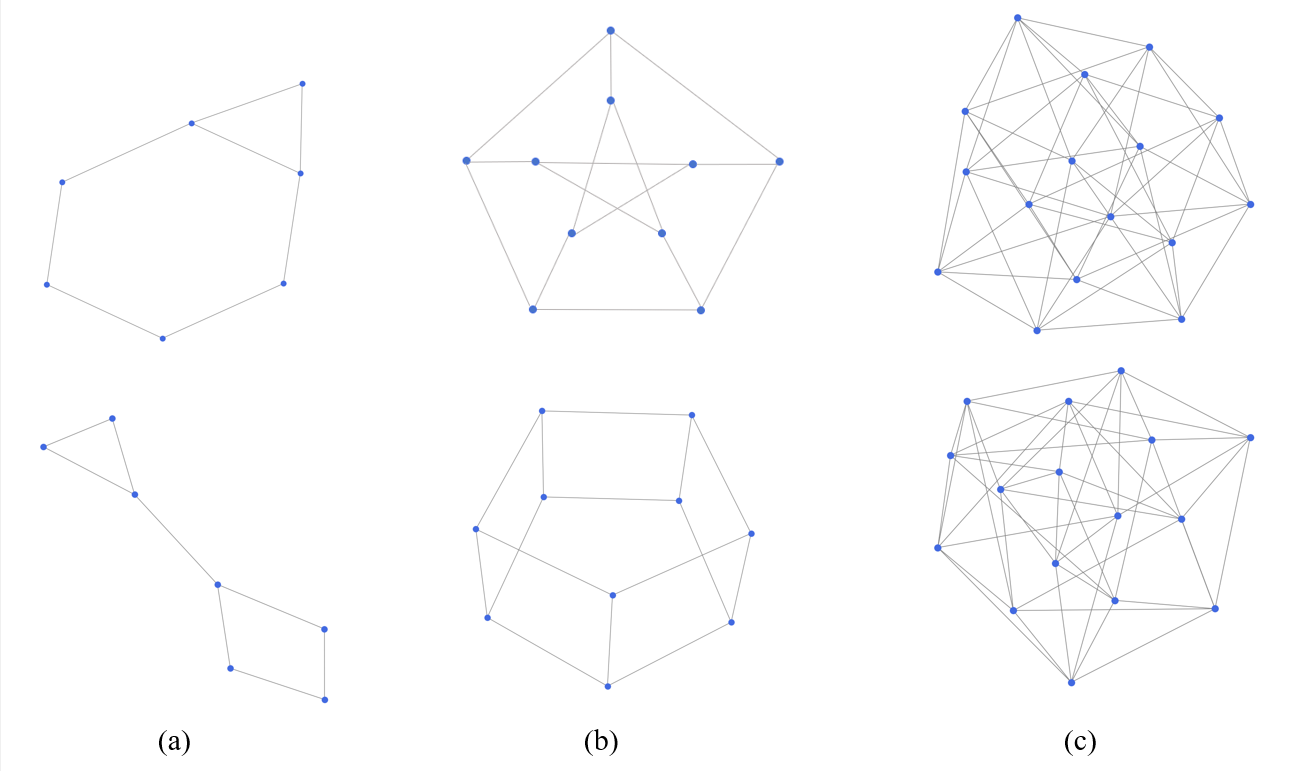}
\caption{Three different types of non-isomorphic graph instances, all with the same degree sequence. (a)is a general graph instance with $N=7$ vertices, its degree sequence is $[2,3,2,2,2,2,3]$. (b) is a regular graph instance with $N=10$ vertices, and its degree sequence is $[3,3,3,3,3,3,3,3,3,3]$. (c) is a strongly regular graph instance with $N=16$ vertices, its degree sequence is $[6,6,6,6,6,6,6,6,6,6,6,6,6,6,6,6]$.} \label{fig7}
\end{figure*}

The local energy $E_{loc}$ is calculated using the sample configuration, repeated $M$ times, and the overall energy expectation of the system is estimated by the sample energy expectation value $\bar{E}_{loc}$(equation\eqref{eq:19}). We transform the graph isomorphism problem into the problem of solving the ground state of the quantum system $\hat{H}$, meaning that we aim to minimize $E_{global}$ by continuously adjusting the parameters of the variational wave function. This adjustment ensures that the probability amplitude of the eigenstate $|\vec{s}\rangle$, which contains the target solution information, increases gradually. The approximation to the ground state is achieved by tuning the wave function parameters $\Omega=\{\vec{a},\vec{b},W\}$, corresponding to the variational part of the VMC algorithm. This approach provides a systematic method for approximating the ground state energy and obtaining the optimal mapping for the graph isomorphism problem through the VMC method.

\begin{figure*}     
\centering  
\includegraphics[width=4.0in]{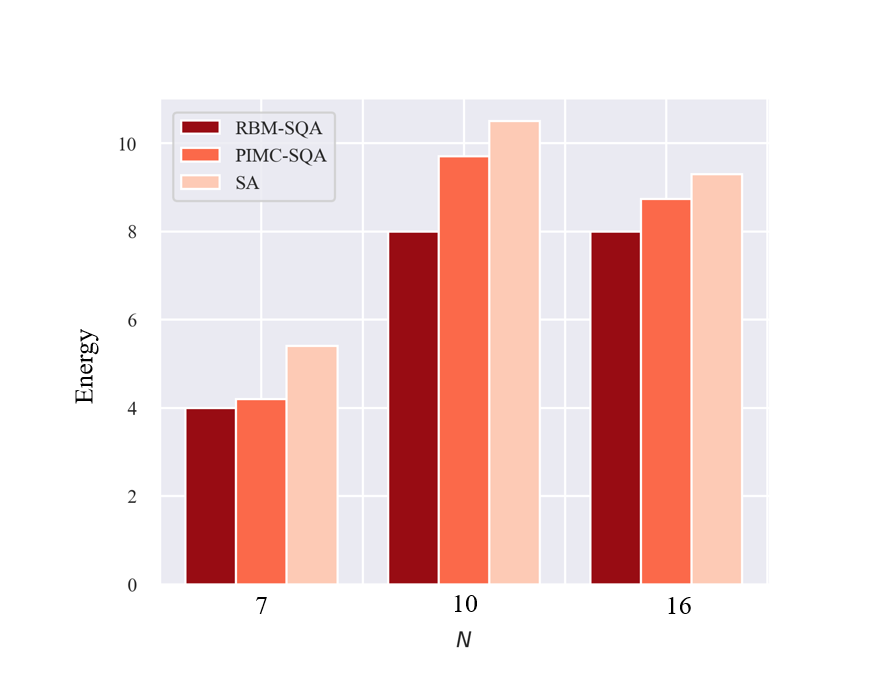}
\caption{Average minimum system energies achieved by SA, PIMC-based SQA, and RBM-based SQA across the different non-isomorphic graphs in Fig.\ref{fig7}.} \label{fig12}
\end{figure*}

As the scale of the problem solution space increases, the Hilbert space dimension of the quantum system becomes very large. The sampling component of the VMC algorithm helps mitigate this issue by leveraging sampling techniques based on relevant statistical principles \cite{gros1989physics}. To ensure statistical precision in estimation, the sample size must be large enough. However, increasing the sampling scale significantly increases computational time and memory consumption. Therefore, a balance between computational precision and resource expenditure is maintained to ensure that sufficiently accurate results can be obtained within reasonable computational resource limits. Sampling introduces statistical uncertainty into the algorithm, and according to the zero-variance principle of VMC, when the system’s wave function $|\psi\rangle$ approaches the true ground-state wave function, the expected variance of the sample energy tends to $0$. In this case, $\bar{E}_{loc}=E_{global}= 0$\cite{mohamed2020monte}. 

Therefore, this is used as an indicator in the experiments to determine whether the system has converged to the ground state. When the variance reaches zero, all local energy values become identical, indicating that the system has reached a stable state.  Additionally, when the energy expectation of the samples no longer changes with successive sampling and optimization steps, it signifies that the parameter update step size is approaching zero and the model has converged. At this point, the sample expectation value provides the best approximation to the energy of the ground state of the system. This also suggests that, in the probability distribution of the solution space after training, the probability amplitude of the target solution is typically much higher than that of non-target solutions, which leads to a significant improvement in the hit rate of the algorithm once the model converges. Although, in practice, the wave function in the VMC algorithm is usually an approximation and it is challenging to achieve the ideal situation of zero variance, this principle provides an optimization criterion for the algorithm. Meanwhile, reducing the variance of the sample energy expectation and minimizing the energy expectation value work together in the gradient estimation of the variational part of the VMC algorithm, collaboratively guiding the adjustment of the wave function parameters. This enables a more precise estimation of the true ground state of the system. Fig.\ref{fig11} illustrates the complete process of this algorithm.

We take the $4$ vertexes isomorphic graph in Fig.\ref{fig3} as an example to check how the target distribution of samples changes with the optimization of $\Omega$. According to the instance encoding shown in Fig.\ref{fig3}, the length of the mapping solution vector is $6$, resulting in a solution space of size $2^6$. The inspection result is Fig.\ref{fig6}. The set of initial parameters $\Omega$ is randomly initialized from a normal distribution with a mean of $0$ and a standard deviation of $0.01$ by default.

In the first iteration, the algorithm produces sample solutions $[0,1,1,0,1,1]$, $[1,1,0,0,1,0]$, and $[0,1,1,1,0,1]$, 
from which it estimates an expected energy of
$\bar E_{loc}=6.25$. To drive the network toward lower-energy configurations, the parameter set $\Omega$ is updated using stochastic gradient descent (SGD) combined with stochastic reconfiguration (SR), effectively moving 
$\Omega$ in the direction of $\nabla_\Omega \bar E_{loc}$
This cycle of sampling, energy estimation, and parameter update repeats at each iteration. As the VMC procedure progresses, the probability distribution over bit-strings gradually concentrates on low-energy (high-quality) mappings. By the 5th iteration, sampled configurations already begin to cluster around the true solution. Continuing this process for 30 iterations brings the variational wave function parameters close to convergence: the probability of the target mapping reaches 0.983, effectively suppressing all incorrect assignments. Finally, a final sampling from the converged NQS yields $\vec{x}=[x_{AB},x_{BA},x_{CC},x_{CD},x_{DC},x_{DD}]$ obtained is $[1,1,0,1,1,0]$, which corresponds to the isomorphism $\{A\leftrightarrow B,B\leftrightarrow A,C\leftrightarrow D,D\leftrightarrow C\}$.
This demonstrates that, through iterative energy minimization and variational parameter updates, the algorithm successfully uncovers the correct graph mapping.
\begin{figure*}    
\centering  
\includegraphics[width=7.0in]{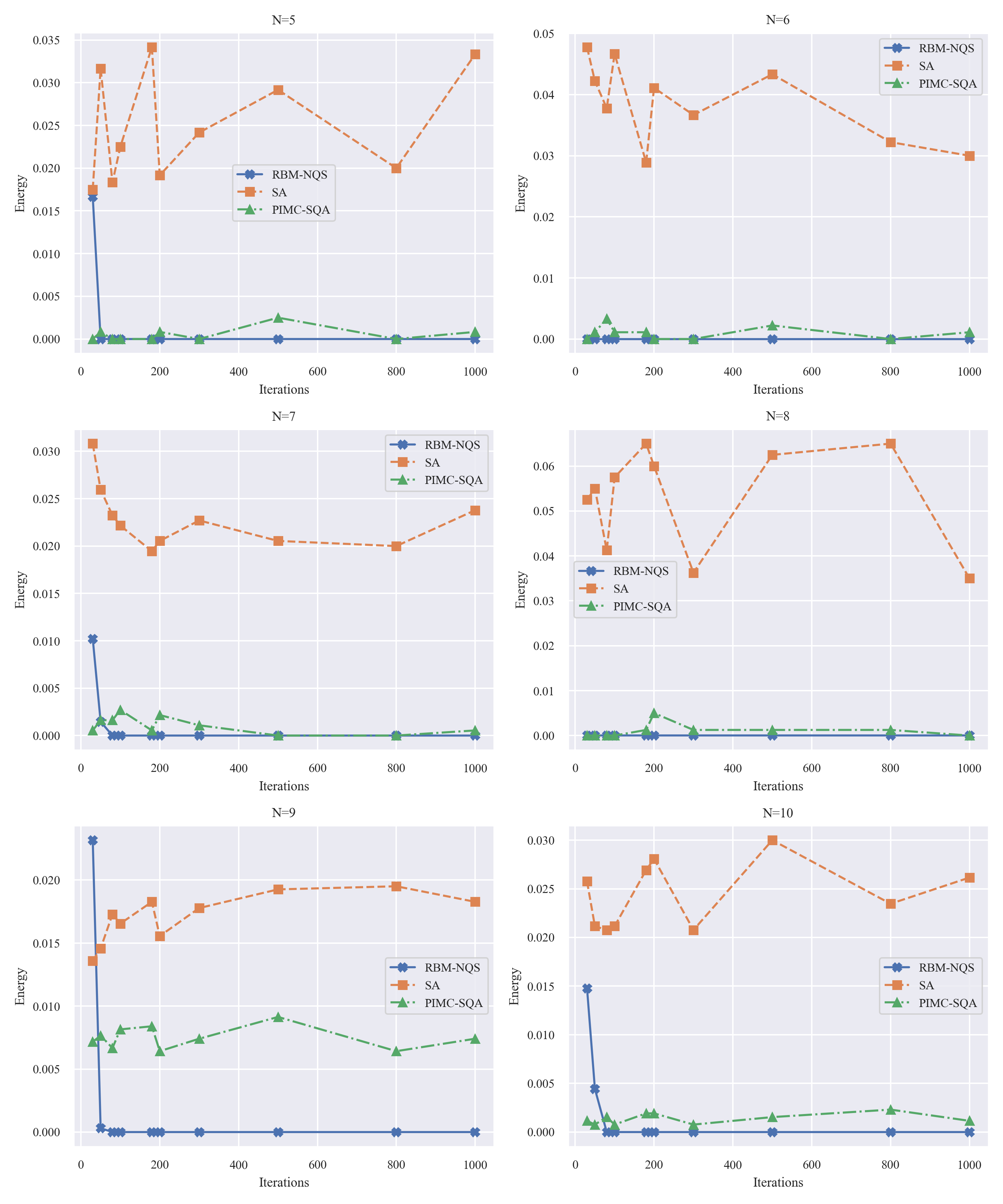}
\caption{The average residual energy of graph instances $N=5,6,7,8,9,10$ evolves over iterations. Although the NQS method starts at a higher energy level than SA and SQA, it achieves a steeper energy decline via variational optimization and settles into a more stable ground state.} \label{fig8}
\end{figure*}

\section{Numerical Validation}\label{Numerical Validation}

\begin{figure*}     
\centering  
\includegraphics[width=7in]{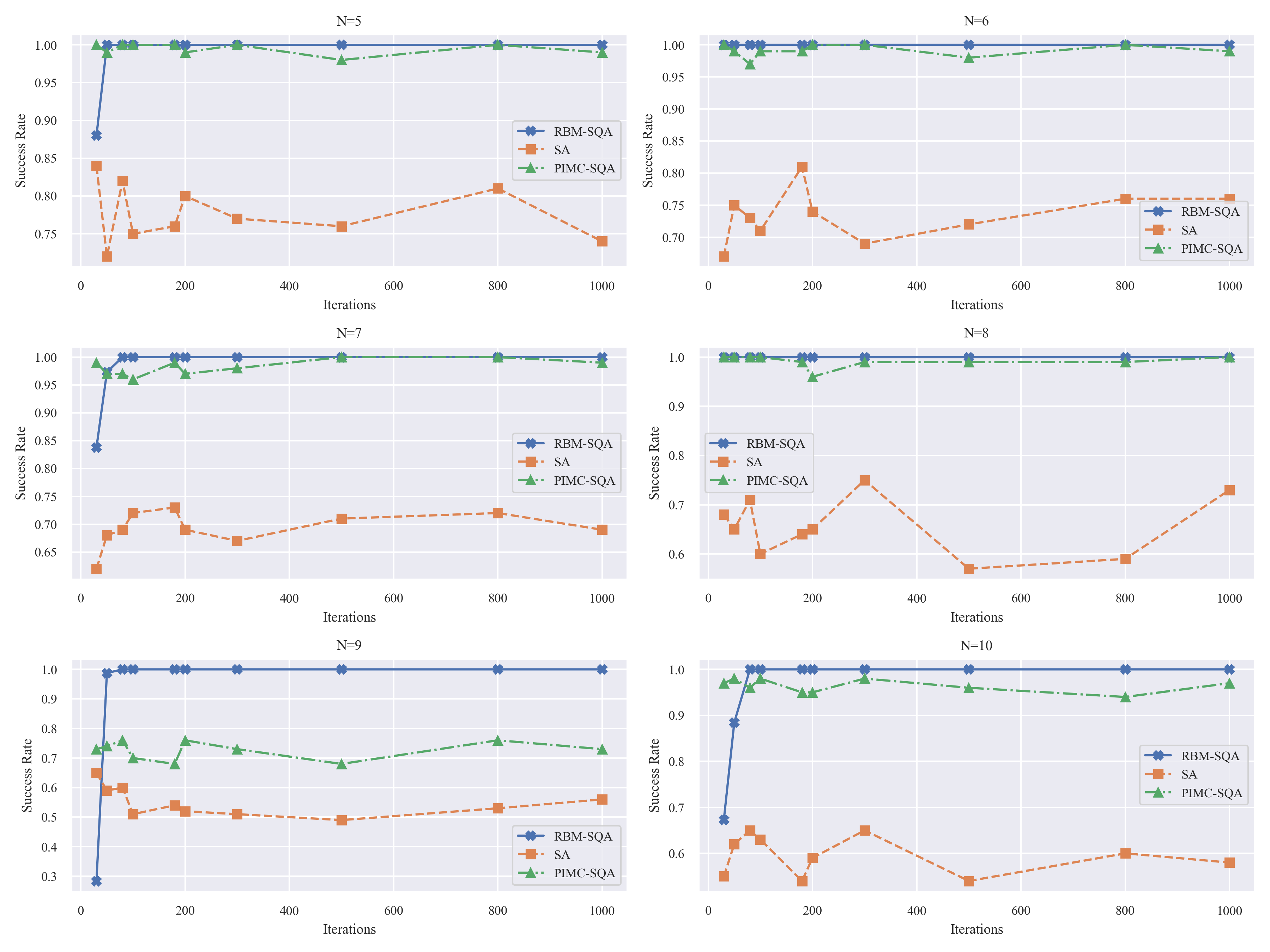}
\caption{The hit rate of SA, PIMC-based SQA, and RBM-based SQA as the number of iterations changes.} \label{fig9}
\end{figure*}

In this paper, we evaluate the performance of our NQS- and VMC-based algorithm against classical simulated annealing (SA) and simulated quantum annealing (SQA) methods, measuring both solution accuracy and runtime when solving the graph isomorphism problem. The experimental setup primarily focuses on testing these algorithms on pairs of isomorphic graphs, assessing each algorithm’s ability to recover the correct vertex mapping. Additionally, experiments were conducted on three different types of non-isomorphic graphs to further demonstrate the capability of the NQS and VMC-based algorithm in distinguishing non-isomorphic graphs. Before presenting these comparative results, we first review the SA and SQA algorithms and outline their parameter settings.

\subsection{Implementation Details of SA and SQA Algorithms}

  SA is a classical probabilistic optimization technique inspired by thermal annealing. At high temperatures, the system has a probability of accepting suboptimal solutions, which in principle helps escape from local minima rather than the other classical heuristic methods. As a typical classical heuristic algorithm, SA performance will be compared with the RBM-based algorithm in this paper. To implement the SA algorithm, we follow these steps: 


\begin{enumerate}
\item Initialization. Start by initializing a solution vector $\vec{x_0}$. Set the initial temperature of the system to $T_0$ and design the objective function $F$ of the graph isomorphism problem, we will refer to it as the energy function later. 
\item Exploring the Solution Space. Select a new solution $\vec{x_1}$ randomly from the neighborhood of the current solution, which is achieved by flipping a random bit in the solution vector.  Compute the energy change $\delta F=F(\vec{x_0})-F(\vec{x_1})$ and update the solution using the Metropolis criterion. If $\Delta E \leq 0$, accept the new solution; if $\Delta E > 0$, accept it with a probability of $P(\Delta E,T)=e^{-\Delta E/T}$. This probabilistic acceptance of non-optimal solutions helps escape local minima. Note that the system temperature remains unchanged in this step. We define $N_{sweep}$ as the number of solution space explorations at a fixed temperature, where $N_{sweep}$ new solution vectors are generated by flipping bits.
\item Annealing. Gradually reduce the system temperature according to the set temperature schedule, transitioning from $T_0$ to $T_1$.
\item Repeat the solution space exploration and annealing steps until the system temperature reaches the final value $T_{final}$. The number of temperature drops is denoted by $N_{annealing}$, which, together with $T_0$ and $T_{final}$, defines the annealing schedule of the SA algorithm. The process ends when the temperature reaches $T_{final}$, outputting the historical optimal solution found during the exploration.
\end{enumerate}

Unlike SA, SQA leverages Trotter slicing to emulate quantum tunneling, allowing it to surmount energy barriers more effectively. In true quantum annealing, adiabatic evolution of the system’s Hamiltonian enables the state to “tunnel” through narrow, high barriers that would trap purely thermal algorithms.  SQA partially replicates this advantage and tends to outperform SA in many scenarios, although it lacks genuine quantum coherence and entanglement. It relies on the path-integral method (PIMC) to statistically mimic quantum tunneling, based on imaginary time evolution. This approach transforms the quantum system’s time-dependent evolution into a thermodynamic equilibrium problem of an equivalent classical system \cite{herman1982path}.

For a given Hamiltonian $H$, the imaginary time evolution operator is defined as $\widehat{U}(\beta) = e^{-\beta H}$, where $\beta = \frac{1}{T}$ is the inverse of the system temperature. For any initial state $|\psi_0\rangle$, the system can evolve toward the ground state $|\psi_f\rangle$ through imaginary time evolution, satisfying:
\begin{equation}
\lim_{\beta \to \infty}\hat U(\beta)|\psi_0\rangle=|\psi_f\rangle
\label{eq:20}
\end{equation}

Assuming the ground state energy is $E_f$,  the ground state $|\psi_f\rangle$ satisfies the eigenvalue equation:
\begin{equation}
H|\psi_f\rangle=E_f|\psi_f\rangle
\label{eq:21}
\end{equation}

According to the theory of thermodynamic equilibrium, the probability of a quantum state under temperature $T$ satisfies:
\begin{equation}
P(|\psi\rangle,T)=\frac{1}{Z}e^{-\beta\langle\psi|H|\psi\rangle}=\frac{1}{Z}\langle\psi|e^{-\beta H}|\psi\rangle=\frac{1}{Z}\langle\psi|\widehat{U}(\beta)|\psi\rangle,
\label{eq:22}
\end{equation}
where $Z$ is the partition function, $Z=\sum_{|\psi\rangle} e^{-\beta \langle \psi|H|\psi\rangle}$. When the system is in thermal equilibrium and the temperature is low, ignoring the degeneracy of the ground state, the probability of the ground state is typically close to $1$.

To facilitate the calculation of the imaginary time evolution $\widehat{U}(\beta)|\psi\rangle$,the Trotter-Suzuki decomposition is commonly employed.
This decomposes the imaginary time evolution operator into the product of local operators. In doing so, we map the original problem Hamiltonian into a classical system consisting of  $\tau$  coupled copies of the original system, known as Trotter slices. Taking the Hamiltonian of the Ising model as an example, after mapping, it is
\begin{equation}
H(t)=-\sum_{k=1}^{\tau}(\sum_{(i,j)}J_ij \sigma_i^k\sigma_j^k+J(t)\sum_{i=1}^N\sigma_i^k\sigma_i^{k+1}),
\label{eq:23}
\end{equation}
where $\tau$ is the number of Trotter slices, $\sigma_i^k$ is the $i$-th spin on the $k$-th slice, $J(t)=-\frac{\tau T}{2}ln(tanh(\frac{\Gamma (t)}{\tau T}))$ is the coupling function between slices, which is the coupling along the imaginary time dimension, and $J_{ij}$ describes the coupling in the original two-dimensional direction of the Ising model, where $\Gamma(t)=\Gamma_0(t-1)$ is a linear function of time, representing the transverse field strength of the quantum system, time $t\in[0,1]$.On this basis, the Metropolis-Hastings Monte Carlo algorithm is used to explore the solution space. Each exploration is called a Monte Carlo Step (MCS), and each MCS consists of one local flip and one global flip.

\begin{itemize}
\item Local movement: A randomly selected spin in a single slice is flipped, and the energy difference is calculated. The change is accepted or rejected using the Metropolis-Hastings criterion, allowing the system to escape from local optima. 
\item Global flip: All spins of the same qubit across all slices are flipped simultaneously, and the new solution is evaluated based on the energy difference.
\end{itemize}
SQA reduces system energy by gradually decreasing the transverse field strength. Similar to SA, it explores the solution space through multiple MCS at each transverse field strength. When $\Gamma$ decreases from $\Gamma_0$ to $\Gamma_{final}$ or the system energy reaches $0$, annealing is completed. The slice with the lowest energy is then selected as the result. 
In the early stages, even if a local optimum is surrounded by a potential barrier, the system still has a probability of tunneling into a better solution space, rather than gradually climbing over the energy barrier like in SA. As time progresses, the transverse field strength decreases, and quantum tunneling behavior gradually fades, causing the system's behavior to transition from quantum tunneling to classical optimization, eventually stabilizing at a (possibly local) optimal solution. 
Although SQA has strong capabilities in avoiding local optima, its efficiency is limited by the simulation method and computational cost. For certain problems, the tunneling rate may be very small, especially when the energy barrier is wide and high, making the algorithm require a long time. Additionally, tunneling may lead to the selection of non-optimal solutions if the optimal solution is surrounded by regions with complex barrier structures.


Our RBM-based SQA uses RBM to parameterize the quantum system, allowing the optimization process to no longer be confined to a fixed classical energy function. Instead, it can variationally optimize the parameters of the wavefunction, enabling RBM to reshape the distribution of the solution space to some extent. By creating more efficient "virtual landscapes" and using parameterized models to guide the search and improve the distribution, the RBM makes the representation of certain low-energy states more concentrated, thereby optimizing the search process, improving the barrier structure of the landscape, and significantly reducing the probability of getting trapped in local optima. However, SA and SQA do not alter the energy landscape of the objective function itself; the local minima and barriers still exist, but the algorithms use different approaches to cross these barriers.

To compare the three algorithms, we set the exploration count for SQA and SA as $N_{sweep}$ and the cooling steps as $N_{annealing}$. For our RBM-based SQA method, the sampling scale per round is set to $N_{sweep}$, and the optimization iterations to $N_{annealing}$. This ensures all algorithms explore the solution space $N_{annealing}*N_{sweep}$ times, with $N_{sweep}$ increasing with graph and solution space scales. By matching the total number of sampling/optimization operations across all methods, we isolate differences in their intrinsic algorithmic efficiency and accuracy. For SA, we choose an initial temperature $T_0=100$ and a final temperature $T_{final}$ of $0.667$. For SQA, we set the transverse-field strength to decrease from  $\Gamma_0=100$ to $\Gamma_{final}=0.667$, maintaining relative consistency with the SA algorithm. For details on the hyperparameter settings of the three algorithms, see Table~\ref{tab2}.

\begin{table}[b] 
\caption{\label{tab2}%
Hyperparameter settings of the three algorithms.}
\begin{ruledtabular}
\begin{tabular}{ccc}
    SA & PIMC-SQA & RBM-SQA \\
    \hline
    $T_0=100$		&$\Gamma_0=100$  &learning rate $\in[0.01,0.1]$ \\
$T_{final}=0.667$		&$\Gamma_{final}=0.667$  &\\
		&Totter slices$=4$  &\\
  \end{tabular}
\end{ruledtabular} 
\end{table}

\begin{figure*}   
\centering  
\includegraphics[width=5.0in]{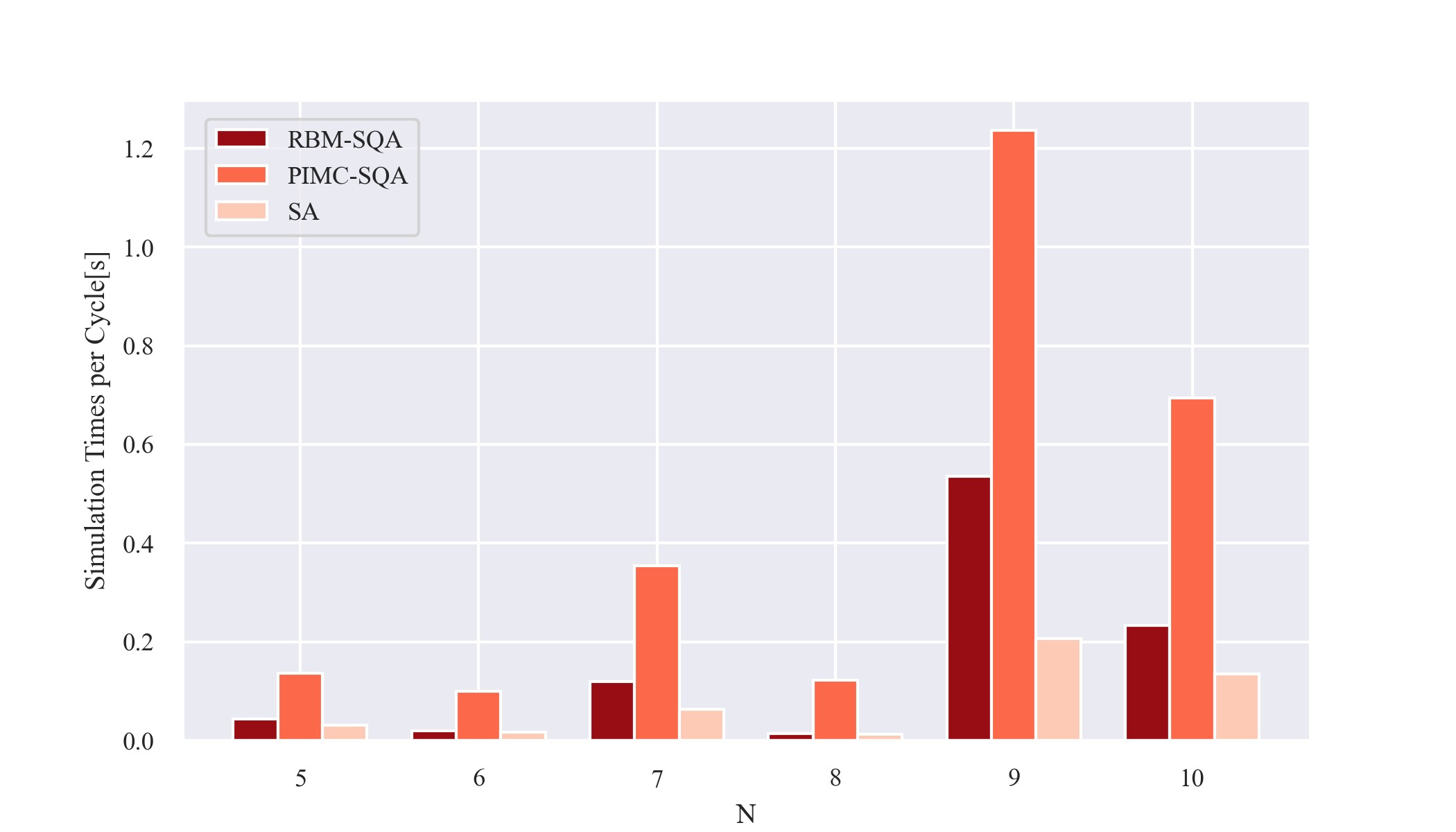}
\caption{A comparison chart of the runtime per iteration in seconds for the three algorithms. RBM-based SQA  typically maintains a median runtime across all isomorphic graph instances in the experiments, while the runtime of the SA and SQA algorithms lies at the two extremes.} \label{fig10}
\end{figure*}

\subsection{Non-isomorphic Graph}

We tested non-isomorphic graphs using RBM-based SQA, demonstrating its ability to distinguish them. Since graphs with different degree sequences can be identified as non-isomorphic during preprocessing, the non-isomorphic graphs tested in this section all share the same degree sequence.

We conducted experiments on several types of graphs (as shown in Fig.\ref{fig7}). A graph $G$ is
$k$ regular if all vertices $v$ satisfy $Deg(v)=k$.For example, in $N=10$ vertex graphs, the Petersen graph and the Pentagonal Prism graph are both $3$ regular due to their high symmetry and $Deg(v)=3$ for all vertices. Strongly regular graphs are classified by parameters $(N,k,\lambda,\mu)$, where graphs with identical parameters belong to the same family. Here, $N$ denotes the number of vertices, $k$ the degree of each vertex, $\lambda$ the number of common neighbors for adjacent vertices, and $\mu$ the number of common neighbors for non-adjacent vertices. For instance, in $N=16$ vertex graphs, $G_1$ and $G_2$ are non-isomorphic but belong to the same family $(16,6,2,2)$. The Hilbert space sizes of these graph instances are $2^{29},2^{100},2^{256}$, respectively. In this experiment, the three algorithms similarly partition the vertex set based on the degree sequence of the graph and then encode it to generate the system's solution space. 
Based on the topological structures and degree sequences of the graphs shown in Fig.~\ref{fig7}, the length of the solution vector is calculated using a vertex-mapping encoding scheme, which determines the required number of qubits and the dimensionality of the solution space. Table~\ref{tab4} presents a summary of the number of vertices $N$, the size of the solution space, the number of qubits, and the sampling scale per iteration for three types of non-isomorphic graphs for our RMB-based SQA. The non-isomorphic graph with $N=7$ vertices (degree sequence $[2,3,2,2,2,2,3]$) requires $29$ qubits, leading to a solution space of size $2^{29}$ and adopts a sampling size of $290$ per iteration, approximately ten times the number of qubits. This parameter setting is sufficient to achieve reasonably accurate solutions in this experiment.

\begin{table}[b] 
\caption{\label{tab4}%
The settings of non-isomorphic parameters.}
\begin{ruledtabular}
\begin{tabular}{cccc}
    N & Space Size & Encoding Qubits & Sampling Size \\
    $7$ &   $2^{29}$ & $29$ & $290$ \\
    $10$ &   $2^{100}$ & $100$ & $1000$ \\
    $16$ &   $2^{256}$ & $256$ & $2560$ \\
  \end{tabular}
\end{ruledtabular} 
\end{table}

As the model parameters converge, the probability amplitude distribution of the quantum system dynamically concentrates towards the low-energy states of the Hamiltonian corresponding to the target problem. Spin configurations that align with lower energy expectations have significantly increased solution probabilities, while those that do not align see a corresponding decrease in solution probability. In particular, the system can only reach the theoretical ground state (with a system energy eigenvalue of $0$) when the two graphs are isomorphic. In all non-isomorphic instances, after 100 samples using the RBM-based SQA, the experimental ground state energy consistently stabilizes at integer energy levels $\geq1$. This indicates that there are no valid isomorphic mappings in subsequent samples. To mitigate the risk of misjudgment, this algorithm considers a batch of samples valid only when the energy error is $0$, ensuring that the mathematical form of the energy expectation is a discrete integer value. This fundamentally eliminates the phenomenon of false convergence, where the energy expectation merely approaches $0$.

Fig.\ref{fig12}  illustrates the average energy corresponding to the optimal solutions obtained by the three algorithms, indirectly reflecting the quality of solutions achieved after annealing across 100 samples. A lower expected energy intuitively indicates a higher degree of optimization in the spin configuration for vertex mapping, manifesting as an increased proportion of vertex pairs in candidate solutions that satisfy isomorphism constraints. Notably, even under non-isomorphic conditions, the algorithm can still achieve maximal partial isomorphic coverage. In scenarios involving non-isomorphic graph instances, RBM-based SQA maintains the lowest average energy, outperforming both SQA and SA algorithms.

\subsection{Isomorphic Graph}
We evaluated our RBM-based SQA algorithm on pairs of isomorphic graphs with vertex counts $N=\{4,5,6,7,8,9,10\}$. For each graph size, we recorded the final hit rate and the average per-iteration runtime once the model had converged. Additionally, for each 
$N$, we examined four distinct degree sequences, each may require a different number of encoding qubits and yield a different solution‐space dimensionality.  These data are shown in Table~\ref{tab5}.

\begin{table}[] 
\caption{\label{tab5}%
Experimental results of the RBM-based SQA algorithm on various isomorphic graph instances. The last column,  "Time" represents the average time per iteration.}
\begin{ruledtabular}
\begin{tabular}{ccccc}
    N & Qubits & Space Size & Final Hit Rate &  Time (s) \\
    $4$ & $6$ & $2^{6}$ & $1$ & $0.00438$ \\
        & $8$ & $2^{8}$ & $1$ & $0.00487$ \\
        & $16$ & $2^{16}$ & $1$ & $0.00828$ \\
        & $16$ & $2^{16}$ & $1$ & $0.01013$ \\
    
    $5$ & $9$ & $2^{9}$ & $1$ & $0.00471$ \\
        & $9$ & $2^{9}$ & $1$ & $0.00512$ \\
        & $25$ & $2^{25}$ & $1$ & $0.02736$ \\
        & $25$ & $2^{25}$ & $1$ & $0.04405$ \\
    
    $6$ & $12$ & $2^{12}$ & $1$ & $0.00550$ \\
        & $12$ & $2^{12}$ & $1$ & $0.00614$ \\
        & $14$ & $2^{14}$ & $1$ & $0.00777$ \\
        & $18$ & $2^{18}$ & $1$ & $0.01945$ \\
    
    $7$ & $19$ & $2^{19}$ & $1$ & $0.01227$ \\
        & $21$ & $2^{21}$ & $1$ & $0.01668$ \\
        & $25$ & $2^{25}$ & $1$ & $0.02281$ \\
        & $37$ & $2^{37}$ & $1$ & $0.11903$ \\
    
    $8$ & $16$ & $2^{16}$ & $1$ & $0.01406$ \\
        & $32$ & $2^{32}$ & $1$ & $0.04284$ \\
        & $38$ & $2^{38}$ & $1$ & $0.07155$ \\
        & $40$ & $2^{40}$ & $1$ & $0.07658$ \\
    
    $9$ & $33$ & $2^{33}$ & $1$ & $0.04150$ \\
        & $45$ & $2^{45}$ & $1$ & $0.10126$ \\
        & $53$ & $2^{53}$ & $1$ & $0.18448$ \\
        & $81$ & $2^{81}$ & $1$ & $0.53516$ \\
    
    $10$ & $34$ & $2^{34}$ & $1$ & $0.04739$\\
         & $44$ & $2^{44}$ & $1$ & $0.08890$\\
         & $50$ & $2^{50}$ & $1$ & $0.13730$\\
         & $52$ & $2^{52}$ & $1$ & $0.23352$\\
  \end{tabular}
\end{ruledtabular} 
\end{table}

To further analyze the specific details of the RBM-based SQA algorithm during its execution, we selected one representative isomorphic graph pair for each  $N = \{5, 6, 7, 8, 9, 10\}$ and benchmarked it against classical SA and PIMC-based SQA. Fig.\ref{fig8} shows the experimental results in each graph instance with $N=\{5,6,7,8,9,10\}$ vertices. Both of these annealing algorithms explore the solution space by simulating a gradual cooling process to find the global optimum. Three algorithms partition the vertex set based on the degree sequence of the graph, and then encode and generate the corresponding solution space. Table \ref{tab3} lists, for each 
$
N$, the degree sequences, solution-space dimensions, qubit counts, and per-iteration sample sizes (set to ten times the qubit count, mirroring Table \ref{tab4}).

\begin{table}[b] 
\caption{\label{tab3}%
The settings of isomorphic graph parameters.}
\begin{ruledtabular}
\begin{tabular}{ccccc}
    N & Degree Sequence & Space Size & Qubits & Sampling Size \\
    $5$ & $[2,2,2,2,2]$ & $2^{25}$ & $25$ & $250$ \\
    $6$ & $[1,2,2,3,2,2]$ & $2^{18}$ & $18$ & $180$ \\
    $7$ & $[3,3,3,3,3,3,6]$ & $2^{37}$ & $37$ & $370$ \\
    $8$ & $[5,4,5,4,3,2,1,4]$ & $2^{16}$ & $16$ & $160$ \\
    $9$ & $[8,8,8,8,8,8,8,8,8]$ & $2^{81}$ & $81$ & $810$ \\
    $10$ & $[7,7,8,8,7,8,8,8,8]$ & $2^{52}$ & $52$ & $520$ \\
  \end{tabular}
\end{ruledtabular} 
\end{table}

Experimental results show that the RBM-based SQA achieves a faster reduction in system energy, converging more quickly to the theoretical minimum.  While SA and PIMC-based SQA algorithms control the search direction by lowering the system temperature and transverse field strength, the RBM-based SQA  accounts for the overall distribution of the system. This enables the RBM-based SQA to make more precise adjustments, resulting in a stable and consistent decrease in energy, ultimately maintaining the theoretical global minimum. 

We also observe that RBM–SQA often starts with higher initial energy values. 
This is partly due to the inherent randomness introduced during the sampling step.
Furthermore, in the RBM-based SQA  method, the energy expectations $\bar{E}_{loc}$ are calculated after randomly initializing neural network parameters and batch sampling using the M-H sampling method. Unlike SA and SQA algorithms that retain only a single optimal solution vector from batch sampling in the solution space, the RBM-based SQA  accounts for the overall distribution.
Although this distribution-based estimate can inflate early energy readings, it does not impair convergence; instead, it enables more accurate, distribution-aware adjustments that accelerate the overall energy decline compared to the other two algorithms.

After training, we sample the estimated ground state and compute the hit rate by counting exact-mapping solutions.  For comparison, we ran SA and PIMC-based SQA algorithms multiple times and calculated their success rates by counting the number of times the final solutions matched the target solution. Fig.\ref{fig9} shows that our algorithm initially has a lower success rate, but it rapidly increases with more sampling iterations and stabilizes at 1, with lower fluctuations and higher overall hit rates than SA and PIMC-based SQA. Thus, our method, when handling rugged and glassy optimization landscapes, achieves faster sampling and a higher probability of finding the target solution compared to algorithms based on classical simulated annealing principles. While simulated annealing can explore a broader state space by "hopping" through thermal fluctuations or escaping local minima via simulated quantum tunneling, it cannot alter the sampling probability. Consequently, its sampling dynamics are slower, and the probability of hitting the target solution is lower in complex optimization landscapes.

Meanwhile, we compared the runtime of these three algorithms to complement the difficulty of accurately estimating the complexity of heuristic algorithms. Fig.\ref{fig10} shows that the SA algorithm has the fastest runtime but only moderate success rates, while the PIMC-based SQA algorithm improves accuracy at the cost of significantly increased runtime. The RBM-based SQA  maintains optimal success rates with runtime falling between the two, within an acceptable range. Additionally, the runtime does not increase with the number of graph vertices, which is related to the problem encoding method. The number of vertices represents only part of the graph information; more complex structures and additional edges often lead to longer convergence times for the algorithm.

\section{Conclusions}\label{Conclusions}

As an emerging approximate quantum system model, NQS has great potential in quantum many-body physics. We encode the graph isomorphism problem as a QUBO problem, map it to the Ising Hamiltonian, and complete the equivalent transformation to the ground state solution of the quantum many-body system. The method was simulated on a classical computer with RBM network quantum
state and compared with MIPC-based simulated annealing algorithms. 
Moreover, the RBM-based SQA replaces the random search principle in the solution space of the SA and SQA algorithms with a variational approach. 
Through the parameterized optimization of the neural network, it can more precisely adjust the distribution of the solution space, thereby more effectively searching for the global optimum, reducing the likelihood of getting trapped in local optima, and improving the accuracy of the algorithm. 
Experimental results show that the average optimal solution hit rate of the proposed algorithm improves by 31.84$\%$ and 4.38$\%$ compared to the PIMC-based SQA and SA algorithms, respectively. Additionally, compared to PIMC-based SQA, our approach achieves an average reduction of 67.44$\%$ in computational time per iteration, demonstrating significant advantages in both accuracy and efficiency for solving graph-related problems.

On this basis, one of our future research directions is to find ways to further reduce the size of the problem solution space. Due to the encoding method, the quantum system corresponding to the correct solution in the problem solution space is often presented in the form of Dicke states. Therefore, we began to think about whether the solution space can be restricted to Dicke state, thereby directly excluding candidate solutions that do not obey the bijective constraints. However, due to the characteristics of the VMC method, even if the initial state is a Dicke state, there is no guarantee that the subsequent state still obeys the Dicke state constraint in the sampling flip operation. Therefore, how to combine Dicke states and variational methods to effectively reduce the size of the solution space needs further research and thinking. In addition, the development of new quantum algorithms, such as quantum neural networks and quantum enhanced optimization methods, will also be an important part of this research direction. These efforts will open up new paths for the future combination of quantum computing and machine learning, thereby achieving breakthrough progress in the field of graph isomorphism problems.

\begin{acknowledgments}
This research was supported by the National Nature Science Foundation of China (Grants No. 62101600), State Key Lab of Processors, Institute of Computing Technology, CAS under Grant No. CLQ202404, the Beijing Natural Science Foundation (Grant No. 4252006), National Natural Science Foundation of China (Grant No. 62301454), Fundamental Research Funds for the Central Universities (Grant No. SWU-KQ22049), and the Natural Science Foundation of Chongqing, China (Grant No. CSTB2023NSCQ-MSX0739).
\end{acknowledgments}


\bibliography{apssamp}

\end{document}